\documentclass[reprint,amsmath,amssymb,prb,aps,showkeys, superscriptaddress]{revtex4-2}

\usepackage{graphicx,xcolor}
\usepackage{dcolumn}
\usepackage{bm, multirow}
\usepackage{mathrsfs}
\usepackage[colorlinks=true, allcolors = blue]{hyperref}
\usepackage{braket}
\usepackage{booktabs}
\usepackage{soul}
\usepackage[utf8]{inputenc}

\begin{document}


\title{Crystal Electric Field Analysis on the Magnetic Properties of  \\  Ferromagnetic CeRu$_2$Ge$_2$ Single Crystal}
\author{Shovan Dan}
\email{dan.shovan@gmail.com}
\affiliation{Department of Condensed Matter Physics and Materials Science, Tata Institute of Fundamental Research, Colaba, Mumbai 400005, India}

\author{Suman Nandi}
\affiliation{Department of Condensed Matter Physics and Materials Science, Tata Institute of Fundamental Research, Colaba, Mumbai 400005, India}

\author{Gourav Dwari}
\affiliation{Department of Condensed Matter Physics and Materials Science, Tata Institute of Fundamental Research, Colaba, Mumbai 400005, India}

\author{Bishal Baran Maity}
\affiliation{Department of Condensed Matter Physics and Materials Science, Tata Institute of Fundamental Research, Colaba, Mumbai 400005, India}
\author{Bhagyashree A Chalke}
\affiliation{Department of Condensed Matter Physics and Materials Science, Tata Institute of Fundamental Research, Colaba, Mumbai 400005, India}
\author{Ruta Kulkarni}
\affiliation{Department of Condensed Matter Physics and Materials Science, Tata Institute of Fundamental Research, Colaba, Mumbai 400005, India}

\author{P. D. Babu}
\affiliation{UGC DAE Consortium for Scientific Research Mumbai Centre, 246 C, Common Facility Building BARC Campus, Mumbai 400085, India}

\author{Arumugam Thamizhavel}
\email{thamizh@tifr.res.in }
\affiliation{Department of Condensed Matter Physics and Materials Science, Tata Institute of Fundamental Research, Colaba, Mumbai 400005, India}

\date{\today}

\begin{abstract}

We report a detailed anisotropic studies on the magnetic and transport properties of a CeRu$_2$Ge$_2$ single crystal. A clear ferromagnetic transition is observed at $T_{\rm C} = 7.5$~K in the magnetic susceptibility, electrical resistivity and heat capacity measurements.  The magnetoresistance (MR) is positive at low temperature within the magnetically ordered state. We confirmed the presence of a magnon excitation gap in the resistivity, MR and the low temperature heat capacity data. The pronounced anisotropy observed in the magnetic susceptibility and magnetization along the [100] and [001] directions is quantitatively explained within the crystal electric field (CEF) framework using the point-charge model. The analysis reveals a CEF level scheme comprising three doublets, with a ground state and two excited states at 491 and 671~K, that was further confirmed from the Schottky effect in the magnetic part of the heat capacity data.  Unlike most Kondo lattice ferromagnets of Ce compounds, where the magnetic easy axis deviates from the expected  CEF ground state, CeRu$_2$Ge$_2$ exhibits magnetization, which is fully consistent with the CEF derived ground state wave function, with the easy axis of magnetization aligned along the [001] direction.
\end{abstract}

\maketitle

\section{Introduction}

Cerium-based intermetallic compounds constitute a prototypical class of materials for exploring strong electron correlations, in which the competition among Kondo screening, magnetic exchange interactions, and crystal electric field (CEF) effects gives rise to a rich variety of quantum phenomena~\cite{Onuki2018,Radousky2000,Stewart1984,Cornut1972}. The single localized $4f^1$ electron of the Ce$^{3+}$ ion, characterized by a $J = 5/2$ multiplet, is extremely sensitive to the local electrostatic environment. This sensitivity leads to a lifting of the degenerate ($2J+1$) levels into 3 doublets or a doublet and a quartet depending on the site symmetry of the Ce-atom. Far from being a mere spectroscopic feature, this CEF splitting provides a microscopic basis for understanding several macroscopic properties, including magnetic anisotropy, specific heat, and paramagnetic susceptibility. Moreover, CEF excitations strongly influence charge transport, as inelastic scattering from excited CEF levels leaves distinct fingerprints in the electrical resistivity and magnetoresistance. Consequently, a comprehensive CEF analysis that integrates both thermodynamic and transport measurements is essential for unraveling the underlying physics of these systems.

\begin{figure*}
	\centering
	\includegraphics[width=0.95\textwidth]{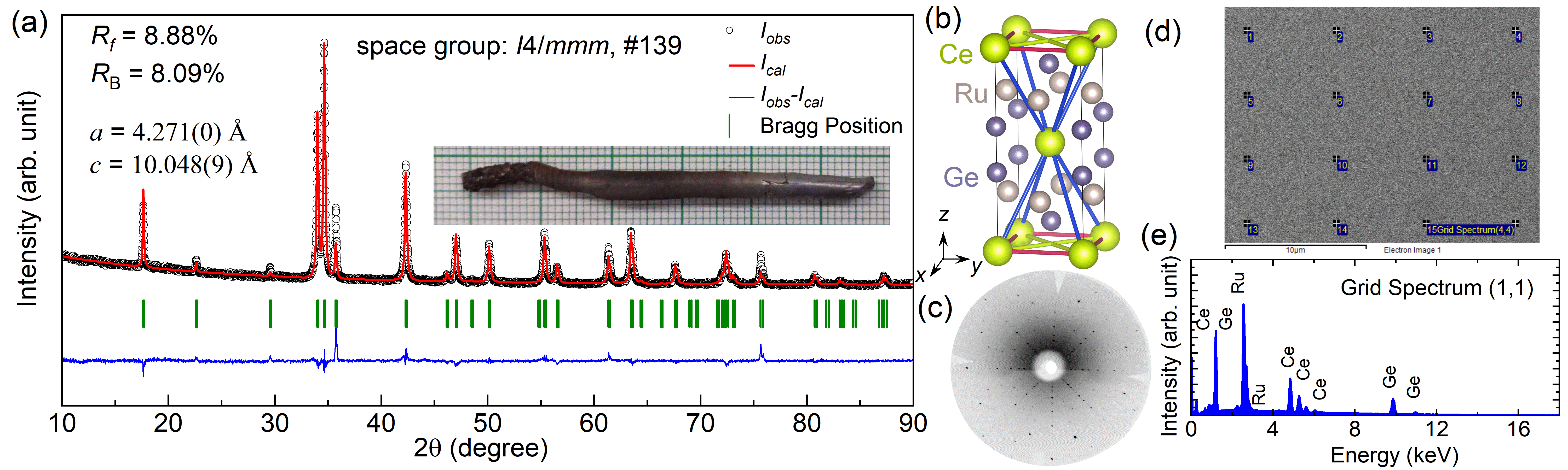}
	\caption{(a) Room temperature powder X-ray diffraction pattern of CeRu$_2$Ge$_2$, as grown crystal is shown in the inset. (b) Crystal structure (using VESTA package~\cite{Momma2011}), nearest neighours, second nearest neighours and third nearest neighours are shown in lime, blue and ruby colors, respectively.  (c) The back-reflected Laue pattern on the (001) plane. (d) Angle Selective Backscattered (ASB) image showing 16 point spatial array grid from which EDS was obtained. (e) EDS spectrum.}
	\label{fig:XRD}
\end{figure*}
Among cerium-based materials, the Ce-based 1-2-2 family is one of the most extensively investigated material classes owing to its remarkable diversity of structural, magnetic, and electronic properties. In particular, the CeM$_2$Ge$_2$ (M = Cu, Ru, Au, Rh, Ag) series crystallizing in the famous ThCr$_2$Si$_2$ type tetragonal crystal structure has attracted sustained interest over the past several decades due to its diverse magnetic ground states~\cite{Thompson1994, Loidl1992, King1991, Kobayashi1998, Godart1987}. For instance, CeRu$_2$Ge$_2$ exhibits a ferromagnetic ground state, whereas CeAu$_2$Ge$_2$ orders antiferromagnetically. In contrast, the Cu- and Ag-based compounds display spiral magnetic structures with incommensurate ordering vectors~\cite{Loidl1992}. Although CeRu$_2$Ge$_2$ is primarily ferromagnetic, reports have suggested the presence of an additional antiferromagnetic phase in the vicinity of the Curie temperature ($T_{\rm C}$), attributed to a fragile and unstable magnetic ground state~\cite{Loidl1992}. Pressure-dependent resistivity measurements have further revealed a quantum phase transition near 10 GPa, marked by a transformation from ferromagnetic to antiferromagnetic order, followed by the emergence of Fermi-liquid behavior~\cite{Sullow1999,Thompson1994,King1991,Kobayashi1998}. It is noteworthy that the isostructural compound CeRu$_2$Si$_2$, which does not exhibit a long-range magnetic order, nevertheless shows signatures of proximity to a quantum critical point even at ambient pressure~\cite{Kobayashi1998}. Extensive investigations involving chemical substitution at the Ge site (Si for Ge) or at the Ru site (Os, Pd, or Rh for Ru) have been carried out to systematically tune the unit cell volume and conduction electron density, thereby providing further insight into the interplay between structure and correlated electron behavior~\cite{Godart1987,Godart1986,Pandey2023}. As the CEF also plays a crucial role in determining the magnetic ground state, a detailed study was carried out on several isostructural compounds~\cite{Loidl1992}. In this manuscript, we have made a comprehensive study on a high quality CeRu$_2$Ge$_2$ single crystal grown by the Czochralski method. 

\section{Experimental details}
A single crystal of CeRu$_2$Ge$_2$ has been grown directly from the stoichiometric melt by the Czochralski method in a tetra-arc furnace (Techno Search Corporation, Japan) under an ultrapure argon atmosphere. 
High-purity starting elements of Ce (99.9\%), Ru(99.99\%), and Ge(99.999\%), in the molar ratio 1:2:2, were melted to obtain a polycrystalline ingot of CeRu$_{2}$Ge$_{2}$. 
To ensure homogeneity, the melting process was repeated multiple times. A seed crystal cut from this ingot was inserted into the molten charge and initially pulled at a rate of 50 mm/h. Once a steady-state growth condition was achieved, the pulling rate was reduced to 10 mm/h to grow the single crystal. 

The cross sectional surface and the chemical composition were characterized using a Carl Zeiss Ultra Plus 55 field emission scanning electron microscope (FE-SEM)  and  Oxford Instruments energy-dispersive X-ray spectroscopy (EDS). Compositional uniformity was assessed via EDS detector by acquiring data over a 16-point spatial array ($4 \times 4$ grid) on a cleaved surface on [001] plane.
A portion of the grown crystal was ground into fine powder and examined at room temperature by x-ray diffraction (XRD) to confirm phase purity. The XRD data was analyzed using Rietveld method and  FullProf software package~\cite{Rodriguez1993}.  The single crystal was oriented using the back-scattered Laue diffraction method and subsequently cut along the two principal crystallographic axes, $a$ and $c$, using a wire electric discharge machine. Magnetic measurements were carried out using a superconducting quantum interference device–vibrating sample magnetometer (SQUID-VSM; Quantum Design, USA). A rectangular, bar-shaped specimen with a mass of 28.18 mg was used for the magnetic measurements. Specific heat measurements were performed in a Physical Property Measurement System (PPMS; Quantum Design, USA). The electrical transport properties of the crystal were measured on a rectangular bar-shaped piece, and the electrical contacts were made using a 50~$\mu$m thick gold wire. 

We have also synthesized polycrystalline LaRu$_2$Ge$_2$ by arc melting method and is used as a non-magnetic reference material for the specific heat measurements. 

\section{Results and Discussion}

\subsection{Structural characterization}
\label{ssec:structure}
Figure~\ref{fig:XRD} shows room temperature powder XRD pattern along with the simulated one with the Rietveld refinement. The refinement shows that the compound crystallizes in centrosymmetric $I4/mmm$ (\#139)  space group, where the Wyckoff positions of the Ce, Ru, and Ge atoms are $2a$~(0, 0, 0), $4d$~(0, 0.5, 0.25), and $4e$~(0, 0, 0.3682), respectively. The structural details are in good agreement with the previous report~\cite{Besnus1991}. The compositional study using the EDS spectra indicates average atomic percentages of Ce, Ru, and Ge to be $19.87(\pm0.15), ~41.99(\pm0.27), ~~\rm{and ~} 38.14(\pm0.38)$, respectively, which are in good agreement with the nominal 1:2:2 stoichiometry. The Ce atoms in the (001) plane form a square network with an interatomic distance $r_1 = 4.271$~\AA. On the other hand the second nearest neighbors and the third nearest neighbors of the Ce atoms are on the [111] and [110] directions with the interatomic distances $r_{2}= 5.864$~\AA,  and $r_{3}= 6.041$~\AA, respectively. 
A well defined Laue diffraction pattern along with the four fold symmetry pattern confirms the good quality of the grown crystal (\textit{cf}. Fig.~\ref{fig:XRD}c). 
\begin{figure*}
	\centering
	\includegraphics[width=1\textwidth]{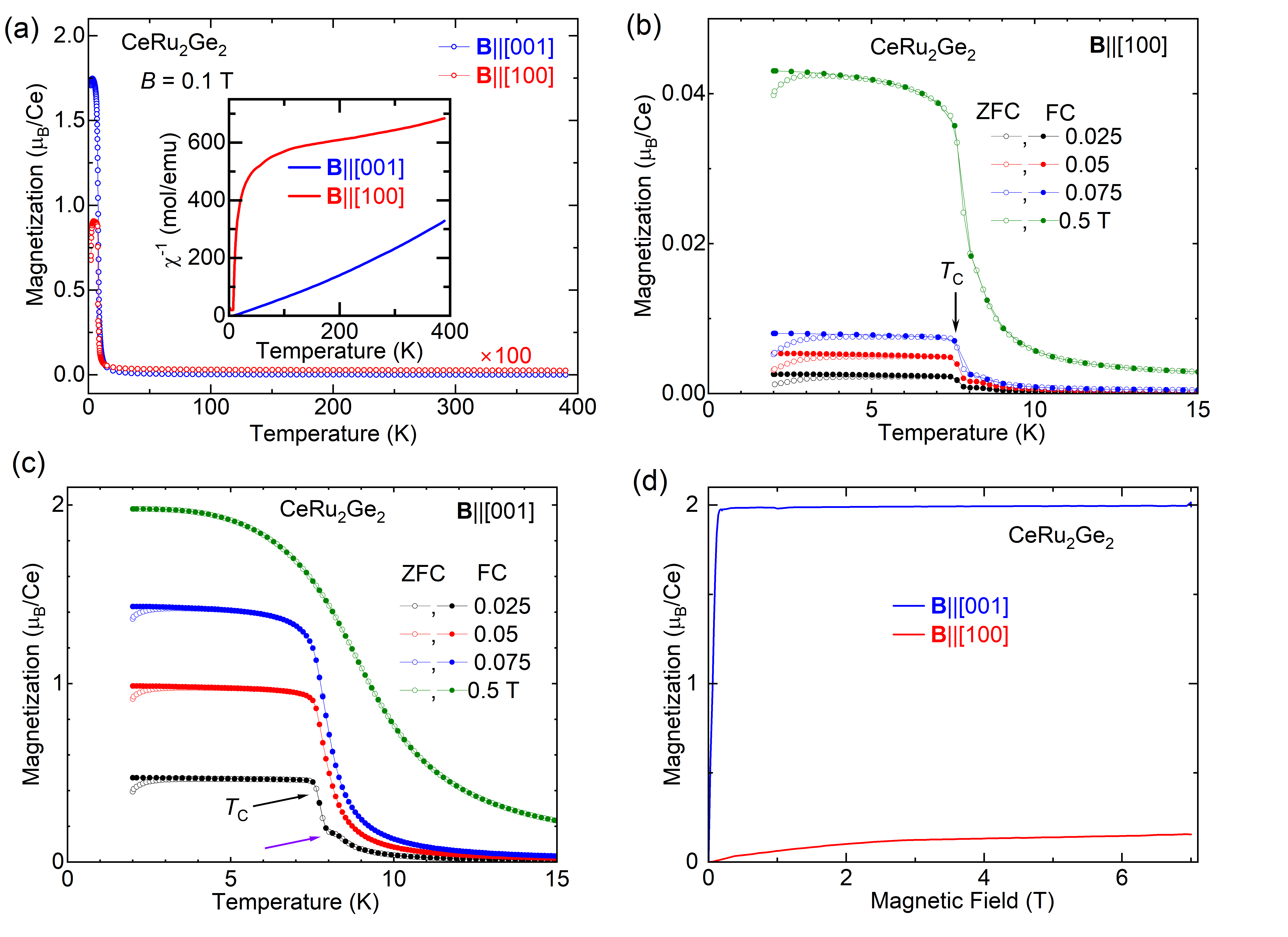}
	\caption{(a) Magnetization as a function of temperature, with magnetic field ($B=0.1$~T) applied on \textbf{B}~$\parallel~[001]$, and \textbf{B}$~\parallel~$[100] directions. Inset shows the inverse susceptibility along both the directions. Magnetization at different applied magnetic field along (b) \textbf{B}~$\parallel~$[001], and (c) \textbf{B}$~\parallel~$[100] directions. The magenta arrow indicates $T_{\rm N}$. (d) Isothermal magnetization  at $T= 2$~K along  \textbf{B}$~\parallel~$[001] and \textbf{B}$~\parallel~$[100] directions.}
	
	\label{fig:M}
\end{figure*}
\subsection{Magnetization}
\label{ssec:magnetization}
\begin{figure*}
	\centering
	\includegraphics[width=1\textwidth]{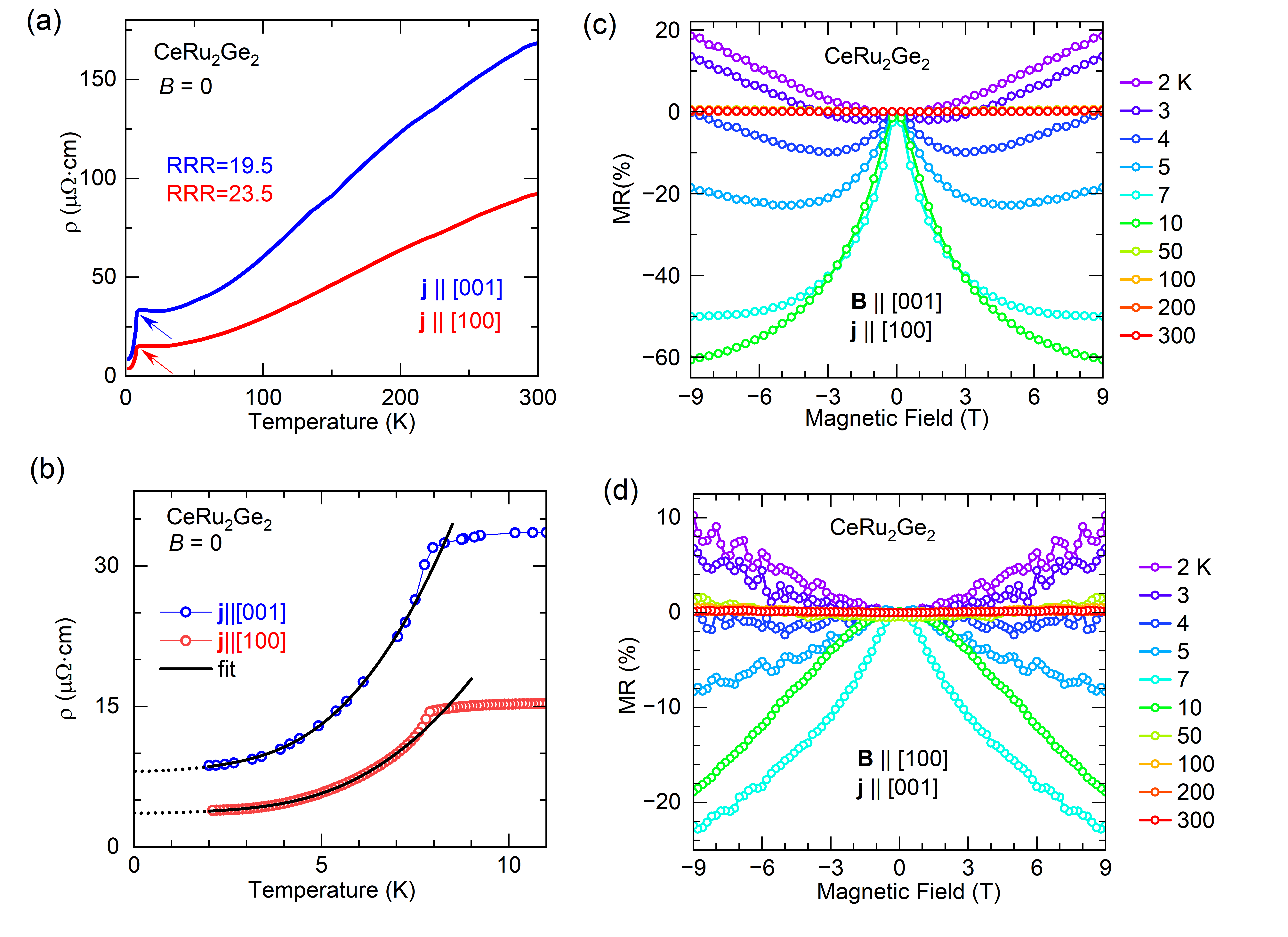}
	\caption{(a) $\rho(T)$ along \textbf{j}$~\parallel~$[001] and \textbf{j}$~\parallel~$[100] direction. The arrows indicate $T_{\rm C}$. (b) Low temperature resistivity along with the magnon-gap fitting. Magnetoresistance  at different temperatures along (c) \textbf{B}$~\parallel~[001]$  and, (d) \textbf{B}$~\parallel~[100]$ directions.}
	\label{fig:Transport}
\end{figure*}

The magnetization~($M$) as a function of temperature~($T$) when the magnetic field ($B= 0.1$~T) is applied along [001] and [100] direction is shown in Fig.~\ref{fig:M}(a). The inverse susceptibility is also plotted in the inset of Fig.~\ref{fig:M}(a)  (the magnetization data of \textbf{B}~$\parallel~[100]$ is multiplied by 100 for clarity. The low field $M(T)$ along \textbf{B}~$\parallel~[001]$  and \textbf{B}$~\parallel~[100]$ is shown in  Fig.~\ref{fig:M}(b,  c).  We have also measured magnetization at different applied magnetic fields, in the temperature region 2--15~K, in both zero field cooled~(ZFC) and field cooled~(FC) conditions. At low field ($B=0.025$~T) along [001] direction, the compound shows a pronounced ferromagnetic ordering at $T_{\rm C}= 7.5$~K. A weak but discernible kink is also observed at 8.2~K (estimated from the first-order temperature derivative of the $M(T)$ data measured at $B=0.025$~T), which disappears at high magnetic field.  Notably, certain allotropic forms of cerium oxides also exhibit antiferromagnetic ordering around 8.5~K. Earlier it was mentioned that the compound shows a second or even a third magnetic ordering close to the $T_{\rm C}$, and the secondary peaks were labeled as antiferromagnetic ordering~\cite{Besnus1991,Rietschel1988}. In the present case, the antiferromagnetic ordering is very weak along [100] direction even at low magnetic field and at high magnetic fields it vanishes similar to that of the [001] direction.  It was also claimed that the second or third peak appears only on the poly-crystalline sample and not in the single crystals. 
A thermomagnetic irreversibility has been observed at low temperature which is much pronounced along \textbf{B}~$\parallel$~[100] direction. The origin of such effect might arise due to the multiple sources and require further investigations.  Magnetic easy and hard directions are clearly evidenced in the isothermal magnetization  at $T= 2$~K along  \textbf{B}$~\parallel~$[001] and  [100] directions [Fig.~\ref{fig:M}(d)]. Along the easy axis, the magnetization saturates at $B\sim 0.2$~T, with a saturation magnetic moment of 1.99~$\mu_{\rm B}$/Ce$^{3+}$, which is almost 93\%  of the full (=$g_J J$~=~2.14~$\mu_{\rm B}$) Ce$^{3+}$ moment. The saturation magnetization is very similar to that reported previously~\cite{Thompson1994,Bohm1988}. On the other hand, the magnetization reaches close to 0.18~$\mu_{\rm B}$/Ce$^{3+}$ at $B=7$~T along the hard axis, $[100]$.
\begin{figure*}
	\centering
	\includegraphics[width=1\textwidth]{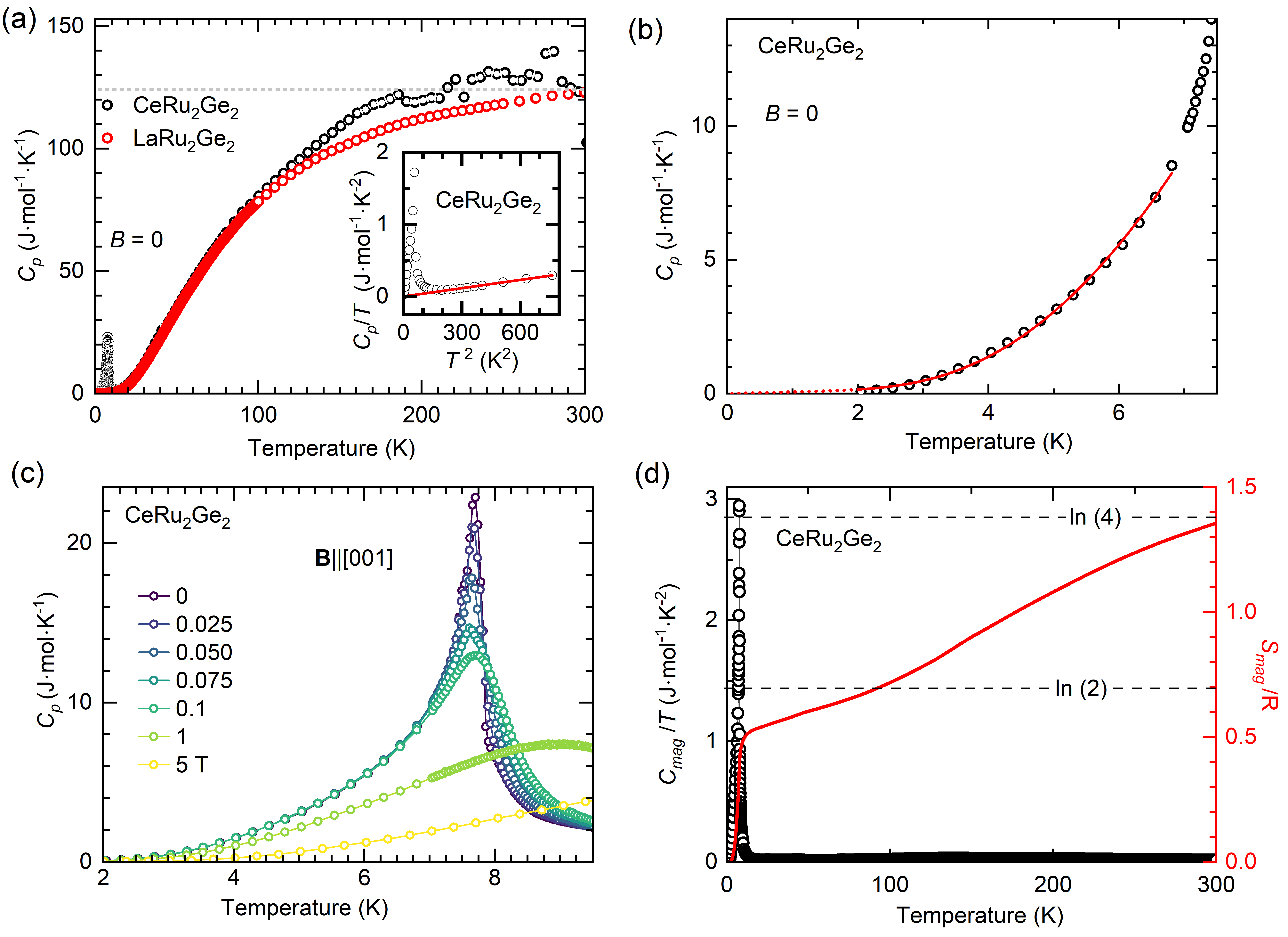}
	\caption{(a) Heat capacity of CeRu$_2$Ge$_2$ and LaRu$_2$Ge$_2$ at $B=0$. The gray dotted line represents $3nR= 124$ J$\cdot$mol$^{-1}\cdot$K$^{-1}$.  The inset shows $C_p/T$ vs. $T^2$ in the low temperature region (b) $C_p$ vs. $T$ below ordering temperature  and fit with magnon excitation energy. (c) $C_p(T)$ at different applied magnetic fields. (d) $C_{\rm mag}/T$, and $S_{\rm mag}/R$ as a function of temperature in the left and right-axis respectively.
	}
	\label{fig:HC}
\end{figure*}
\subsection{Electrical transport}
\label{ssec:transport}
 Anisotropic electrical resistivity along \textbf{j}$~\parallel~$[001] and \textbf{j}$~\parallel~$[100] direction is shown in Fig.~\ref{fig:Transport}(a). The data show an usual dip close to the $T_{\rm C}$.  The temperature dependence of the resistivity for both current directions below the magnetic ordering temperature is shown in Fig.~\ref{fig:Transport}(b).  The low-temperature resistivity is expected to reflect signatures of the magnon excitation spectrum, potentially including the opening of a magnon gap. We have fitted the low temperature resistivity using the magnon gap relation:~\cite{Jobiliong2005,Baumbach2012,Baumbach2012a,Fontes1999,Zhou1996}
\begin{equation}
	\rho (T)= \rho_0+\mathcal{A}T^2+\mathcal{C}T\Delta\left(1+\frac{2T}{\Delta}\right){\rm e}^{-\frac{\Delta}{T}}
	\label{eq:magnon_transport}
\end{equation}
\noindent where $\rho_0$ is the residual resistivity, $\mathcal{A}$ and $\mathcal{C}$ are the fitting coefficients and $\Delta$ is the magnon gap. The fitted  parametes for both the directions are tabulated in Tab.~\ref{tab:magnon}.
\begin{table}[h]
	\caption{Fitting parameters related to magnon scattering}
	\label{tab:magnon}
	\begin{tabular}{cc|c|c|c}\\ \hline \hline 
		& $\rho_0$          & $\mathcal{A}$                    & $\mathcal{C}$                    & $\Delta$ \\
		& ($\mu\Omega\cdot$ cm) & ($\mu\Omega\cdot$ cm$\cdot$ K$^{-2}$) & ($\mu\Omega\cdot$ cm$\cdot$ K$^{-1}$) & (K)   \\ \hline 
		\textbf{j}$~\parallel~$[001] & 8.09          & 0.1255               & 0.4771               & 17.83 \\
		\textbf{j}$~\parallel~$[100] & 3.62          & 0.0528               & 0.2549               & 19.22 \\ \hline
	\end{tabular}
\end{table}
The magnetoresistance [MR= $\frac{\rho(B)-\rho(0)}{\rho(0)}$] for both \textbf{j}$~\parallel~[001]$ and \textbf{j}$~\parallel~[100]$ directions are shown in Fig.~\ref{fig:Transport}(c) and (d). The basic features of the MR for both the directions are similar. A comparable kind of behavior was observed on a few Kondo ferromagnets, \textit{viz.}, CePd$_2$Al$_8$~\cite{Tursina2018}, Ce$_2$Ru$_3$Ge$_5$~\cite{Kamadurai2020}, etc. The upturn with increasing magnetic field is present for both the directions, which gradually follows a downward pattern with increasing temperature.Above the magnetic ordering ($T\geq 50$~K ) the MR becomes very small.
Nevertheless, it is unusual to have an upturn for a ferromagnetic system, below its ordering temperature. Usually, the MR in a typical ferromagnet (ignoring structural disorder, and phononic term as it is measured at constant $T$ ) consists of two components: (1) Lorentz term arising from the cyclotron motion of the electrons, which increases with increasing $B$ and (2) spin-disorder term, which reduces scattering with increasing $B$. In a good metal as the $\omega_c\tau$  term is very small, the Lorentz contribution will be very small~\cite{pippard}. Thus, we have tried to explain the positive MR at $T=2$~K, and its gradual downturn with increasing temperature below $T_{\rm C}$, with the help of magnon excitation energy gap. A qualitative discussion is provided in  Appendix~\ref{app:1}. The MR in the temperature range 2--10~K can be explained using the relative strength of the magnon and the spin disorder contribution. While approaching $T_{\rm C}$, the spin contribution dominates and the overall MR shows a decreasing pattern.

\subsection{Specific heat}
\label{ssec:HC}
The specific heat of both CeRu$_2$Ge$_2$ and LaRu$_2$Ge$_2$ is shown in Fig.~\ref{fig:HC}(a). From the linear fit [inset of Fig.~\ref{fig:HC}(a)] of the $C_p/T$ vs. $T^2$ in the low temperature region, using the relation: $C_p/T=\gamma+\beta T^2$, just above the mgnetic ordering temperature, we have estimated the Sommerfeld coefficient $\gamma = 1.94$~mJ$\cdot$mol$^{-1}\cdot$K$^{-2}$, which is significantly small. We have estimated the value of $\Theta_{\rm D}= 298$~K, from the value of $\beta=3.821\times 10^{-4}$~J$\cdot$mol$^{-1}\cdot$K$^{-4}$ using the relation: $\Theta_{\rm D}=\left(\frac{12\pi^4}{5\beta}nR\right)^{\frac{1}{3}}$. 
Fig.~\ref{fig:HC}(b) shows the enlarged part of the low temperature $C_p$ that contains useful information about the magnon gap. We have tried to fit that using the relation:~\cite{Bohm1988}
\begin{equation}
	C_p(T)= \gamma T+\beta_m T^{\frac{3}{2}} {\rm e}^{-\frac{\Delta}{T}}
	\label{eq:cp_magnon}
\end{equation}
\noindent where $\Delta$ is the energy gap in the magnon spectrum. The fitting yields $\Delta = 11$~K, which is close to the value of estimated by B\"{o}hm \textit{et.al.}~\cite{Bohm1988}. The authors showed that the gap is robust with magnetic field upto $B=4$~T. 
The same report also estimated $\gamma = 20$~mJ mol$^{-1}$K$^{-2}$ and $\beta_m = 2.6$~J mol$^{-1}$K$^{-2.5}$. In our analysis, we obtained $\gamma = 64$~mJ mol$^{-1}$ K$^{-2}$ and $\beta_m = 2.21$~J mol$^{-1}$~K$^{-2.5}$. While the estimated $\beta_m$ value is in close agreement with the reported result, the discrepancy in $\gamma$ may arise from the fact that our data is measured only down to 2~K, while in the published data they have measured down to milli-Kelvin temperature range.  It should also be noted that the $\gamma$ value estimated in the paramagnetic region, just above the magnetic ordering temperature, is not identical to the value obtained below the ordering temperature. The magnon gap ($\Delta$) estimated independently from the low-temperature resistivity and specific-heat data are found to be close in magnitude ($\Delta \approx 17.83$~K and $\Delta \approx 11$~K, respectively), confirming the presence of a gapped magnon excitation spectrum in the ordered state. The modest quantitative difference between the two estimates can be understood in terms of the following factors. First, resistivity and specific heat probe the magnon excitations through fundamentally different mechanisms: the resistivity is governed by the scattering of conduction electrons off thermally populated magnons and is preferentially weighted by the low-lying magnon modes that couple most strongly to the itinerant carriers, whereas the specific heat is a thermodynamic probe that integrates over the entire magnon density of states. The two techniques therefore need not sample identical portions of the magnon dispersion, which can lead to differing extracted gap values. Second, and more importantly,  the specific-heat data in the vicinity of the ordering temperature exhibit clear signatures of a mixed first-order/second-order transition, i.e., a partial first-order character superimposed on the dominant second-order transition. This behavior indicates the presence of short-range magnetic correlations and a degree of phase coexistence close to the ordering temperature, along with an additional latent-heat-like contribution to the anomaly. Considering these differing probe sensitivities and the additional complexity introduced by the mixed-order nature of the transition, the agreement between $\Delta$ extracted by two different probes lying within the same order of magnitude is reasonable and supports a consistent picture of a gapped magnon spectrum in this compound.

However, in the present case, [\textit{cf}. Fig.~\ref{fig:HC}(c)] we have seen that although the $C_p(T,B)$ data below 6~K, and 0.1~T collapse onto a single curve, the data above $B>1$~T is not following Eq.~\ref{eq:cp_magnon}. The magnetic part of the specific heat is calculated by subtracting the specific heat of isostructural compound LaRu$_2$Ge$_2$, and shown in the left axis of Fig.~\ref{fig:HC}(d). 
At $T_{\mathrm{C}}$, the measured $C_{\mathrm{mag}}(T)$ is slightly larger than the mean-field expectation of 12.5~J\,mol$^{-1}$\,K$^{-1}$, corroborating earlier findings reported in Ref.~\cite{Besnus1991}. The magnetic entropy ($C_{\rm mag}$) was calculated using the relation: $S_{\rm mag}=\int_{0}^{T}\frac{C_{\rm mag}}{T}{\rm d}T$. The reduced magnetic entropy $S_{\rm mag}/R$ is plotted in the right axis of Fig.~\ref{fig:HC}(d). For Ce$^{3+}$ with $J= \frac{5}{2}$, the normalized magnetic entropy should saturate at $\ln(2J+1)= \ln6\simeq 1.79$. However, in the present case, the levels are not fully populated, as the CEF split  levels are separated by a large energy range.


\begin{figure*}
	\includegraphics[width=1\linewidth]{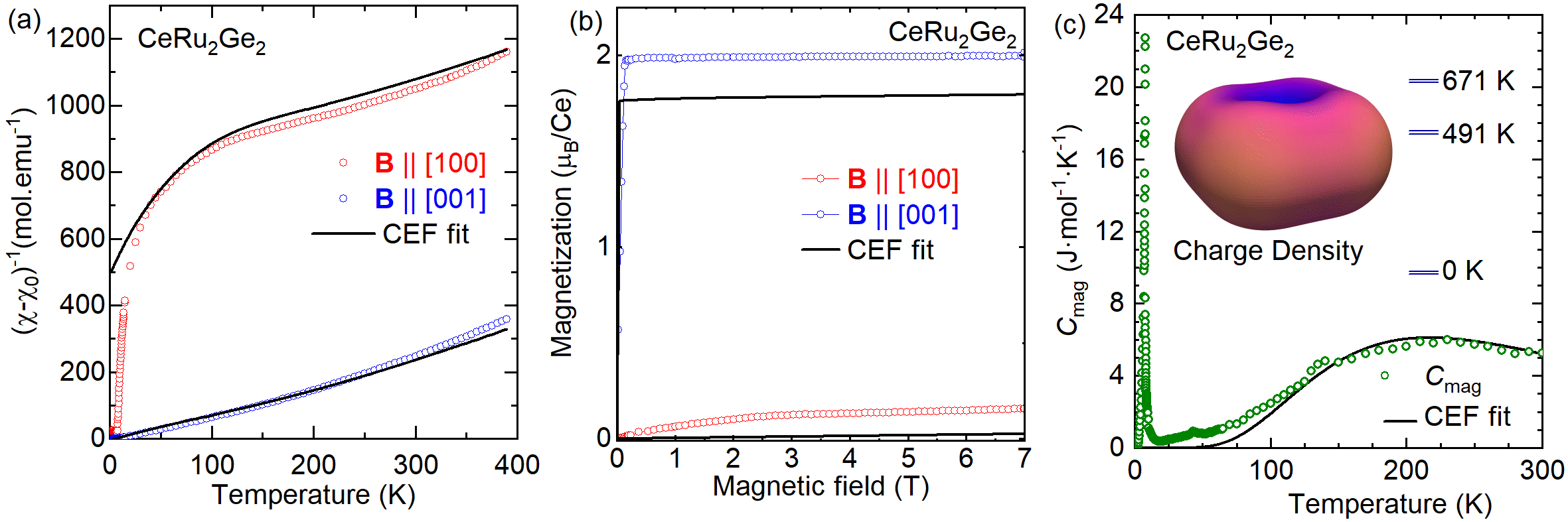}
	\caption{(a) Inverse susceptibility measured at $B = 0.1$~T and (b) Magnetization measured at $T = 2$~K, along with CEF fitting. (c) $C_{\rm mag}$ and CEF fitting using Schottky anomaly. The charge density of the Ce atoms calculated using the obtained CEF parameters is embedded.}
	\label{fig:cef}
\end{figure*}

\subsection{Crystal Field Analysis}
\label{sec:CEF}

We have analysed the magnetocrystalline anisotropy using the point charge model due to the CEF, which lifts the degeneracy of the $2J+1$ ground state of the Ce$^{3+}$ ion. The Hamiltonian can be expressed as:
\begin{equation}
	\mathscr{H}= \mathscr{H}_{\mathrm{CEF}}- g_J \mu_B J_i \left( H + \lambda_i M_i \right)
\end{equation}
\noindent where $\mathscr{H}_{\mathrm{CEF}}$ represents the CEF Hamiltonian and two additional contributions arising from the Zeeman interaction and the molecular field. The variables $g_J$ and $\mu_{\rm B}$ correspond to the $\text{Ce}^{3+}$ Landé g-factor and the Bohr magneton, while $J_i$ and $M_i$ represent the Cartesian components of angular momentum and magnetization, respectively. The CEF Hamiltonian for Ce atom (with $S=1/2$ and $L = 3$) in a tetragonal site symmetry is given by~\cite{Stevens1967, Hutchings1964}
\begin{equation}
\mathscr{H}_{\rm CEF} = B^2_0\mathscr{O}^2_0+B^4_0\mathscr{O}^4_0+B^4_4\mathscr{O}^4_4
\end{equation}
\noindent where $B^m_l$ and $\mathscr{O}^m_l$ represents the CEF parameters and the Stevens operator, respectively. Using this relation, we estimated the effects of CEF on the magnetic susceptibility, magnetization, and specific heat data. The CEF susceptibility and magnetization can be fitted using the usual CEF expressions~\cite{PrSi}.

		\label{eq:chi_2}
We have fitted the linear part (340--390~K) of the susceptibility data using modified Curie-Weiss relation:
\begin{equation}
	\chi= \frac{\rm C}{T-\theta_p}+\chi_0
		\label{eq:CW}
\end{equation}
\noindent where $C$ is the Curie constant,  $\theta_p$ is the paramagnetic Weiss temperature and $\chi_0$ is the temperature independent susceptibility. For the CEF analysis of the inverse susceptibility, the temperature‑independent contribution $\chi_{0}$ was subtracted from the measured susceptibility. The analysis was performed assuming an effective moment of 2.54~$\mu_{\rm B}$/Ce, appropriate for a trivalent Ce$^{3+}$ ion.
The obtained $\chi_0$ was subtracted from the experimental data and fitted with the Eq.~\ref{eq:chi_2}. Similarly, the isothermal magnetization data measured at $T=2$~K was fitted using the CEF relation~\cite{PrSi}.

The fitting of experimental data  yields: $B_2^0= -25$~K, $B_4^0= -0.8$~K, $B_4^4= 8.5$~K, $\lambda_x=-460~{\rm emu}\cdot {\rm mol}^{-1}$ and  $\lambda_z=4.1~{\rm emu}\cdot {\rm mol}^{-1}$. The fitted susceptibility and magnetization along both the directions are shown in Fig.~\ref{fig:cef} (a)  and (b). The obtained CEF parameters and wave-functions are  shown in Table~\ref{tab:CEF}. The eigenfunctions shown in Table~\ref{tab:CEF}, reveals that the ground state is a mixture of $\ket{\pm \tfrac{5}{2}}$
and $\ket{\pm \tfrac{3}{2}}$ and the first excited state is a purely $\ket{\pm \tfrac{1}{2}}$ while the second excited state is dominated by the $\ket{\pm \tfrac{3}{2}}$.   Notably, the molecular field parameters obtained for the $[100]$ direction is negative, whereas, the same for the [001] direction is positive. The large CEF splitting and the strongly Ising like ground state doublet, the susceptibility along the $[100]$ direction (hard-axis) is strongly suppressed.  Within the meanfield CEF analysis, this leads to a very large negative molecular field constant $\lambda_{x}$.  This should be regarded as the large anisotropic exchange rather than as a direct measure of huge antiferromagnetic exchange field along the hard axis.  This type of large negative molecular field constant values are also observed in the ferromagnetic CeRh$_6$Ge$_4$~\cite{Shu2021}, which also exhibts a large magnetocrystalline anisotropy.  Furthermore, the negative value of $B^0_2$ supports the experimentally observed easy magnetic direction along $[001]$-direction. It is to be mentioned here that in this series of compounds CeM$_2$Ge$_2$ (M = Au, Ag, Cu and Ru) except for Ag, the other compounds have the easy axis anisotropy~\cite{Loidl1992,Thamizhavel2007}.  Our estimated CEF energy levels are in close agreement with the energy level schemes obtained from inelastic neutron diffraction data~\cite{Loidl1992}. The CEF derived magnetization is slightly less than the observed experimental moment.  This is due to the mixing of the  $\ket{\pm \tfrac{5}{2}}$, $\ket{\pm \tfrac{3}{2}}$ ground state wavefunction.  

Following Table~\ref{tab:CEF}, the ground state wave function can be written as:
\begin{align}
	\vert \psi_a \rangle = 0.931\vert+\tfrac{5}{2}\rangle -0.365\vert-\tfrac{3}{2}\rangle \\
	\vert \psi_b \rangle = 0.931\vert-\tfrac{5}{2}\rangle- 0.365\vert+\tfrac{3}{2}\rangle
\end{align}

\noindent The ground-state magnetic moment of rare-earth intermetallic compounds is primarily governed by the CEF. In Ce and Yb based intermetallic compounds, however, the Kondo effect can also influence the magnetic moment due to the proximity of the $4f$ level to the Fermi energy. In the present case of CeRu$_2$Ge$_2$, the CEF ground state is not a pure $\lvert \pm \tfrac{5}{2} \rangle$ doublet, and consequently the saturation magnetization along the easy axis is slightly reduced from the expected $2.14~\mu_{\rm B}$/Ce for a trivalent Ce$^{3+}$ ion.  Interestingly, several ferromagnetic Kondo lattice compounds including CePd$_2$Ga$_3$, CeAgSb$_2$, and CeRh$_6$Ge$_4$ etc.~\cite{Bauer1993, Takeuchi2003, Bauer1994, Shu2021, Araki2003, Itokazu2025, Matsuoka2015}, exhibit a CEF-derived ground state that is nearly a pure $\lvert \pm \tfrac{1}{2} \rangle$ doublet, yet the ordered moment is larger along the  hard axis~\cite{Hafner2019}. A recent study on Ce$_2$Rh$_3$Ge$_5$, which orders antiferromagneticaly also reports a similar anomaly: although the CEF ground state favors an easy axis along the $c$ direction, inelastic neutron powder diffraction reveals that the ordered moment lies within the $ab$ plane~\cite{Tripathi2026}. These observations suggest that when the Kondo effect is sufficiently strong, hybridization between conduction electrons and localized $4f$ moments can partially delocalize the $4f$ electrons, enhancing the RKKY exchange along the hard direction and stabilizing an ordered moment larger than that expected from the CEF ground state alone~\cite{Kruger2014, Kwasigroch2022}. In CeRu$_2$Ge$_2$, by contrast, no signatures of Kondo hybridization are observed in thermodynamic or transport measurements, and the system therefore adopts a magnetic ground state that closely reflects the CEF-derived doublet without significant renormalization.
The deviation of the CEF derived inverse susceptibility at low temperature for ${\bf B}~\parallel$~[100] may be attributed to the magnetic ordering and domain formation.  This does not affect the validity of the CEF parameters obtained in the paramagnetic state. Moreover, the CEF energy levels align well with the values estimated by B\"{o}hm et.~al. using inelastic neutron scattering~\cite{Bohm1988}.


\begin{table}[h!]
	\centering
	\caption{Energy level scheme and corresponding wave functions, and CEF parameters of the compound $\text{CeRu}_2\text{Ge}_2$}
	\label{tab:CEF}
	\begin{tabular}{lllllll}
	\toprule \\
		$E$(K) & $\ket{+ \tfrac{5}{2}}$ & $\ket{+ \tfrac{3}{2}}$  & $\ket{+ \tfrac{1}{2}}$  & $\ket{- \tfrac{1}{2}}$ & $\ket{- \tfrac{3}{2}}$ & $\ket{- \tfrac{5}{2}}$ \\ \\
		\hline\\
		671 & 0.365 & 0 & 0 & 0 & 0.931 & 0 \\
		
		671 & 0& 0.931  & 0 & 0 & 0& 0.365  \\
		
		491 & 0 & 0 & 0 & 1 & 0 & 0 \\
		
		491& 0 & 0 & 1 & 0 & 0 & 0 \\
		
		0 & 0 & -0.365 & 0 & 0 & 0 & 0.931 \\     
		
		0 & 0.931 & 0 & 0 & 0 & -0.365 & 0 \\
		\hline \\
		\multicolumn{7}{l}{CEF Parameters:}\\ \\
		\multicolumn{7}{l}{$B_2^0= -25$~K, $B_4^0= -0.8$~K, $B_4^4= 8.5$~K}\\
		\multicolumn{7}{l}{$\lambda_x=-460$~emu$\cdot$mol$^{-1}$,  $\lambda_z=4.1$~emu$\cdot$mol$^{-1}$}\\ \\
		\hline
	\end{tabular}
\end{table}

Next, we analysed the magnetic part of the heat capacity $C_{\rm mag}$ obtained after subtracting the lattice heat capacity of LaRu$_2$Ge$_2$.  A broad peak is observed in the $C_{\rm mag}$  centered around 200~K due to the thermal population of the excited levels.  We analysed this Schottky heat capacity~\cite{PrSi} using the energy gaps estimated by fitting the susceptibility and magnetization data. The obtained CEF energy levels from the analysis of magnetic susceptibility and the isothermal magnetization data clearly explains the observed Schottky peak in the heat capacity, justifying the estimated CEF energy splittings. {In a previous report~\cite{felten1987specific} on CeRu$_{2.16}$Ge$_2$, a similar Schottky peak was observed around $\sim 220$ K, with crystal electric field (CEF) energy levels at approximately 500 and 750 K, which are in close agreement with our present results.
The electronic charge density of the Ce atom is calculated using the McPhase software package~\cite{Rotter2004} and shown in Fig.~\ref{fig:cef}(c). 



\section{Conclusion}
In this study, the anisotropic magnetic and transport properties of CeRu$_2$Ge$_2$ single crystals were investigated. In addition to the established ferromagnetic transition at $T_{\rm C} = 7.5$~K, a subtle, field-sensitive antiferromagnetic (AFM) anomaly was identified at $T_{\rm N} =  8.2$~K. This fragile AFM state is easily suppressed by external magnetic fields and is absent in specific heat measurements. 
The magnon gap estimated from the fitting of low temperature resistivity data is 17.83 and 19.22~K, for \textbf{j}$~\parallel~[001]$, and \textbf{j}$~\parallel~[100]$ directions, respectively. The magnon gap estimated from the specific heat data is $\sim11$~K, which is close to the value estimated from the resistivity data.  Additionally, the unusual positive MR observed at 2~K is attributed to the presence of a magnon excitation energy gap. As temperature increases toward $T_{\rm C}$, the MR becomes negative from 5~K onwards as the spin disorder contributions become dominant. The pronounced magnetic anisotropy observed between the $[100]$ and $[001]$ directions was quantitatively analyzed using a point-charge model within the crystal electric field (CEF) framework. This analysis yielded a CEF level scheme consisting of three doublets with excitation energies at 491~K and 671~K, which were further corroborated by Schottky anomalies in the specific heat data.  
CeRu$_2$Ge$_2$ exhibits pronounced magnetocrystalline anisotropy; however, no signatures of strong Kondo interactions or significant quantum fluctuations are observed. Consequently, the magnetization behavior is well described by the CEF ground-state wavefunction. This contrasts with other Kondo ferromagnets, where the hard axis often shows enhanced magnetization that deviates from predictions based on the CEF wavefunctions.
Furthermore, electronic charge density calculations around the Ce sites provided a microscopic basis for the observed easy and hard magnetic axes, successfully linking the material's electronic environment to its macroscopic magnetic behavior.

\appendix
\section{Magnetoresistance due to Magnon}
\label{app:1}
\setcounter{equation}{0}

\renewcommand{\theequation}{A\arabic{equation}}
At constant temperature $T$, the resistivity as a function of magnetic field,  with $\Delta = \Delta(B)$, can be expressed as:

\begin{equation}
     \rho(B) = \underbrace{(\rho_0 + {\mathcal A}T^2)}_{\rho_{\rm c}}
    + \underbrace{\mathcal{C}T\,\bigl(\Delta(B) + 2T\bigr)\,e^{-\Delta(B)/T}}_{f(\Delta(B))}
    \label{eq:A1}
\end{equation}

\noindent where $\rho_c$ is the temperature and field-independent part and $ f(\Delta(B)) = \mathcal{C} T(\Delta + 2T)\,e^{-\Delta/T}$ is field dependent part, so that:
\begin{equation}
    \rho(B) = \rho_{\rm c} + f(\Delta(B)), \qquad \rho(0) = \rho_{\rm c} + f(\Delta_0)
    \label{eq:A2}
\end{equation}

\noindent where $\Delta_0 \equiv \Delta(B=0)$. Therefore, from the definition of MR$(B)$ = $\frac{\rho(B) - \rho(0)}{\rho(0)}$, one can rewrite, 

\begin{equation}
    \text{MR}(B) =  \frac{f(\Delta(B)) - f(\Delta_0)}{\rho_{\rm c} + f(\Delta_0)} = \frac{f(\Delta(B)) - f(\Delta_0)}{\rho(0)}
    \label{eq:A3}
\end{equation}

Taking the derivative of $f$ with respect to $\Delta$ provides

\begin{align}
    \frac{df}{d\Delta} 
    &= -\mathcal{C}\left(\Delta + T\right)e^{-\Delta/T}
    \label{eq:A5}
\end{align}

Thus, to calculate $\frac{d(\text{MR})}{dB}$, we have used

\begin{equation}
    \frac{d(\text{MR})}{dB} = \frac{1}{\rho(0)}\frac{df}{dB}
    = \frac{1}{\rho(0)}\frac{df}{d\Delta}\cdot\frac{d\Delta}{dB}
    \label{eq:A6}
\end{equation}
Further simplification provides:
\begin{equation}
    \frac{d(\text{MR})}{dB} =-
\frac{\mathcal{C}\,(\Delta(B) + T)}{\rho(0)}\,e^{-\Delta(B)/T}\cdot\frac{d\Delta}{dB}
    \label{eq:A7}
\end{equation}

As we have considered $\dfrac{d\Delta(B)}{d B}$ is negative, the values in Eq.~\ref{eq:A7} always return a positive value.
\section*{Acknowledgment}
We acknowledge the support of the Department of Atomic Energy, Government of India, under Project Identification No. RTI 4015 and RTI 4016.

%

\end{document}